\documentclass[twocolumn,prd,groupedaddress,nofootinbib,showpacs]{revtex4}%
\usepackage{amsmath}
\usepackage{latexsym}
\usepackage{amssymb}
\usepackage{graphicx}
\usepackage{dcolumn}
\usepackage{bm}%
\usepackage{amsfonts}
\providecommand{\U}[1]{\protect\rule{.1in}{.1in}}
\input{epsf}
\begin{document}
\title{ Vortex Formation in a $U\left(  1\right)  \times U\left(  1\right)  ^{\prime
}-N=2-D=3\ $Supersymmetric Gauge Model}
\author{C. N. Ferreira}
\affiliation{N\'{u}cleo de Estudos em F\'{\i}sica, Instituto Federal de Educa\c{c}\~{a}o,
Ci\^{e}ncia e Tecnolog\'{\i}a Fluminense, Rua Dr. Siqueira, 273, Campos dos Goytacazes,}
\affiliation{Rio de Janeiro, Brazil, CEP 28030-130}
\author{J.A. Helay\"{e}l Neto and A. A. V. Paredes}
\affiliation{Centro Brasileiro de Pesquisas F\'{\i}sicas, Rua Dr. Xavier Sigaud 150, Urca,}
\affiliation{Rio de Janeiro, Brazil, CEP 22290-180}
\date{\today}

\begin{abstract}
In this work, we study a $U\left(  1\right)  \times U\left(  1\right)
^{\prime}-$ model that results from a dimensional reduction of the N=1-D=4
supersymmetric version of the Cremer-Scherk-Kalb-Ramond model with non-minimal
coupling to matter. Field truncations are not carried out, two Abelian
symmetries coexist and three vector fields are present; two of them are gauge
bosons. Then, by considering the full $N=2-D=3$ supersymmetric model, we study
the mechanism for magnetic vortex formation by means of the Bogomol'nyi
relations, the magnetic flux and the topological charge in the presence of the
two gauge potentials. A short discussion on the relation between our $N=2-D=3$
supersymmetric model and vortices in superfluid films is also presented.

\end{abstract}

\maketitle

\section{Introduction}

The study of symmetries in Physics is of crucial importance as a tool for the
understanding and the description of the Elementary Particles and their
process. The wide symmetry behind a Grand-Unified Theory (GUT) accomodates a
large variety of phenomena in a single model. For instance, we have the
Standard Model of the Elementary Particles (SM), where $SU(3)\otimes
SU(2)\otimes U(1)$ describes three types of interactions. Other important
invariances appear in connection with the Standard Model for Particle Physics.
These symmetries are the Lorentz, CPT\cite{CPT} and Supersymmetry (SUSY)
\cite{Weinberg} invariances. Despite the success of this model, there are
already some important questions for wich the SM has not provided a
satisfactory explanation. One of these problems is the mass generation
mechanism given by the Higgs boson, still lacking experimental evidences.
Nowadays, this is of great motivation for detection in the Large Hadron
Collider (LHC). The mass generation mechanism is based on a spontaneous
symmetry breaking. Sometimes, this breaking is realised by a non-trivial
vacuum configuration, so important in the formation of topological deffects.
In this non-trivial vacuum configuration, there may appear vortex
configurations. The study of vortex configurations in a supersymmetric context
is important due to the fact that supersymmetry (SUSY) is considered a
fundamental symmetry in the early universe, where the vortex
configuration\cite{Vilenkin,Kibble1,Vilenkin1} appears together with the
symmetry breaking in GUT scenarios \cite{Kibble1}.

SUSY is a key ingredient in Superstring Theory, the Minimal Supersymmetric
Standard Model and in connection with Neutrino Physics too. On the other hand,
p-form potentials appear in many supersymmetric models. A 2-form field is
referred to as the Kalb-Ramond field (KR) \cite{KR,Lund}. In 4D, this 2-gauge
form can be related with the real scalar field and can be important for the
study the mediators particles of zero spin\cite{Francia05}. The relation
between a scalar particle and a 2-form gauge potential is important to
understand the spontaneous symmetry breaking carried out by KR field in a
Goldstone Model. KR fields are also important in the study of the vortex
superfluids\cite{Davis88,Davis89,Ferreira08}. Other aspect of the KR field is
the study of its non-Abelian gauge generalization that has not yet found a
good formulation but can be associated with a non-linear chiral sigma-model
\cite{Freedmann81} important to study the interaction between extended objects.

In this work, we wish to investigate the complete $N=2-D=3$ gauge model with a
$U\left(  1\right)  \times U\left(  1\right)  ^{\prime}$ symmetry. Similar
models have recently been studied considering a non-Abelian Chern-Simons term
and generic gauge groups \cite{Gudnason}, but without a KR\ field. In a
previous work \cite{Christiansen:1998gj}, the truncated $N=2-D=3$ model
including the KR field has been considered and the vortex configurations have
been worked out. The truncation consisted in identifying fields that appear
from the dimensional reduction of an $N=1-D=4$ model, as studied in
\cite{Christiansen:1998xf}. Here, we reconsider this model and discuss the
full reduced model with two families of gauge potentials with a mixed
Chern-Simons term and we focus on the analysis of vortex-type solutions in the
presence of the second family of gauge fields. The outline of this paper is as
follows: in Section 2, we present some considerations about the
Cremer-Scherk-Kalb-Ramond (CSKR) model in the supersymmetric N=2-D=3
scenario.\textit{\ }In Section 3, we devote our attention to showing the
ingredients of the vortex magnetic configuration, we study the bosonic part of
our SUSY model, the equations of motion and the critical coupling.\textit{\ }%
In Section 4, we study the Bogomol'nyi equations and the minimal energy
configuration of the vortex. Then, in Section 5, the relation beetwen our
$N=2-D=3$ supersymmetric model with vortices in superfluid films is discussed.
Finally, we draw our General Conclusions in Section 6.

\section{The $N=2-D=3$ SUSY model without truncation}

In this section, we briefly review the $N=2-D=3$ model that results from the
dimensional reduction of the four-dimensinal CSKR
model\cite{Christiansen:1998gj}. This model descends from the $N=1-D=4$ action
that describes QED in the supersymmetric version coupled to the Kalb-Ramond
field in a non-minimal way. This non-minimal coupling is unique. To see this,
consider the pure Kalb-Ramond action coupled to an arbitrary current,
\begin{equation}
S_{K-R}=\int d^{3}x\left\{  -\frac{1}{6}G_{\mu\nu\kappa}G^{\mu\nu\kappa
}+J^{\mu\nu}B_{\mu\nu}\right\}  ,
\end{equation}
where $G_{\mu\nu\kappa}$ is the field-strength 3-form.

In momentum space, the field $B_{\mu\nu}\left(  k\right)  \equiv\widetilde
{B}_{\mu\nu}$ can be expanded as follows:%
\begin{equation}
\widetilde{B}_{\mu\nu}=\alpha k^{\mu}k^{\nu}+\beta_{I}k^{\mu}e_{I}^{\nu
}+\gamma_{I}\overline{k}^{\mu}e_{I}^{\nu}+\delta_{IJ}e_{I}^{\mu}e_{J}^{\nu},
\end{equation}
where the basis vectors are taken as below:
\begin{align}
k^{\mu}  & =\left(  k^{0},\overrightarrow{k}\right)  ;\ \overline{k}^{\mu
}=\left(  k^{0},-\overrightarrow{k}\right)  ;\ \label{bb}\\
e_{I}^{\mu}  & =\left(  0,\overrightarrow{e}_{I}\right)  ;\ \overrightarrow
{e}_{I}\cdot\overrightarrow{k}=0,\text{com }I=1,2.\nonumber
\end{align}

With the help of the gauge symmetry for $B_{\mu\nu}$, $\widetilde{B}_{\mu\nu}$
can be shown to acquire the form%
\begin{equation}
\widetilde{B}_{\mu\nu}=\delta_{IJ}e_{I}^{\mu}e_{J}^{\nu}.
\end{equation}

So, the equations of motion in momentum space read as:%
\begin{align}
k^{2}\delta_{IJ}e_{I}^{i}e_{J}^{j}  & =\widetilde{J}^{ij},\\
n\varepsilon^{ijk}k_{k}  & =\widetilde{J}^{ij},\label{1}%
\end{align}
where $n=k^{2}\delta_{IJ}$.

\bigskip Equation (\ref{1}) ensures that the current coupled to the
Kalb-Ramond field is actually a topological current with the form:%
\begin{equation}
J^{\mu\nu}=\varepsilon^{\mu\nu\kappa\lambda}\partial_{\kappa}j_{\lambda}.
\end{equation}

This result denies the possiblity of writting down a symmetry group associated
with the conservation of $J_{\mu\nu}$. In other words, the Yang-Mills version
of the Kalb-Ramond model is not possible and this is actually shown as a no-go
result in the work of \cite{Henneaux:1997mf}.

\bigskip Back to the $N=2-D=3$ model \cite{Becker:1995sp}, we write down the
gauge-field sector of the bosonic action in components as:%

\begin{align}
S_{gauge}  & =\int d^{3}x\{-{\frac{1}{4}}F_{\mu\nu}F^{\mu\nu}%
+\,2\,m\,\varepsilon^{\mu\nu\alpha}A_{\mu}\partial_{\nu}B_{\alpha}\nonumber\\
& -{\frac{1}{2}}G_{\mu\nu}G^{\mu\nu}\},\label{CS}%
\end{align}
where the index $\mu=0,1,2$, with $F_{\mu\nu}=\partial_{\mu}A_{\nu}%
-\partial_{\nu}A_{\mu}$ being the electromagnetic field-strength. $B_{\mu}$ is
the vector given by the reduction of the 4-dimensional Kalb-Ramond field,
$B^{3\mu}$, with a corresponding field-strength $G_{\mu\nu}=\partial_{\mu
}B_{\nu}-\partial_{\nu}B_{\mu}$. Another vector field, the dual of $B^{\mu\nu
}$ in $3D$, comes out which is defined by $B^{\mu\nu}=\varepsilon^{\mu\nu\rho
}Z_{\rho}$. Having in mind that, in $\left(  1+2\right)  D$, the Kalb-Ramond
field-strength may be written as a scalar,%
\begin{equation}
G_{\mu\nu\kappa}=S\varepsilon_{\mu\nu\kappa},\label{bc}%
\end{equation}
then%

\begin{equation}
\partial_{\mu}Z^{\mu}=\frac{1}{2}\varepsilon^{\mu\nu\kappa}\partial_{\mu
}B_{\nu\kappa}=S.
\end{equation}
However, from the free field equations and the gauge transformation $Z_{\mu
}^{\prime}=Z_{\mu}+\varepsilon_{\mu\nu\kappa}\partial^{\nu}\xi^{\kappa}$, $S$
is shown to be a constant, so that $B_{\mu\nu}$ does not correspond to a
physical degree of freedom, unless it interacts with other fields.

The part of the $N=2-D=3$ action involving the scalars is written as follows:%

\begin{gather}
S_{scalar}=\int d^{3}x\{e^{-2gM}\nabla_{\mu}\varphi(\nabla^{\mu}\varphi
)^{\ast}+P\left(  \varphi\right)  \partial_{\mu}M\partial^{\mu}M\nonumber\\
+{\frac{1}{2}}\partial_{\mu}N\partial^{\mu}N\,+\,2mN\left(  \partial_{\mu
}Z^{\mu}\right) \\
\ \ \ \ \ \ \ \ \ -\,g^{2}(\partial_{\mu}Z^{\mu})^{2}|\varphi|^{2}%
e^{-2gM}+\,(\partial_{\mu}Z^{\mu})^{2}\},\nonumber
\end{gather}
where $P\left(  \varphi\right)  =1-g^{2}|\varphi|^{2}e^{-2gM}$. The covariant
derivative, $\nabla_{\mu},$ is given by%

\begin{equation}
\nabla_{\mu}\varphi=(\partial_{\mu}+ihA_{\mu}+igG_{\mu})\varphi\text{.}%
\end{equation}

$M$ and $N$ are real scalars. The dual fields, $F_{\mu}$ and $G_{\mu},$ are
given by:%

\begin{equation}
F_{\mu}={\frac{1}{2}}\epsilon_{\mu\nu\kappa}F^{\nu\kappa};\ \ \ \ \ \ G_{\mu
}={\frac{1}{2}}\epsilon_{\mu\nu\kappa}G^{\nu\kappa}.
\end{equation}

Adopting the parametrisations $\phi=e^{-gM}\varphi$ and $\partial_{\mu}Z^{\mu
}=S$, we write down the remaining piece of the bosonic action, where the
auxiliary field, $\Delta$, is present and from which we can extract the
potential of the model. We denote it by $S_{U}$ and it is given by%

\begin{align}
S_{U}=\int d^{3}x  & \{hN|\phi|^{2}+\,2h\Delta|\phi|^{2}+\nonumber\\
& +2\Delta^{2}-4mM\Delta+\eta\Delta\}.
\end{align}

The equation of motion for the auxiliary field yields%

\begin{equation}
\Delta=mM-{\frac{h}{2}}|\phi|^{2}-{\frac{\eta}{4}.}%
\end{equation}

Once it is eliminated, the potential for the physical scalars takes the form below:%

\begin{align}
U  & ={\frac{h^{2}}{2}}\left(  |\phi|^{2}-{\frac{2m}{h}}M-v^{2}\right)
^{2}+\nonumber\\
& -\left(  h^{2}N^{2}+g^{2}S^{2}\right)  |\phi|^{2}-2SmN-S^{2},
\end{align}
where $\nu^{2}=\frac{\eta}{-2h}$. Once this potential has been built up, we
are ready to discuss the symmetry-breaking pattern that yields the vortex formation.

\section{\bigskip Critical coupling and field equations}

The equations of motion for the fields involved in our Lagrangian density are
given below:
\begin{gather}
\partial_{\nu}\left[  \left(  1-g^{2}|\phi|^{2}\right)  \left(  \partial_{\mu
}Z^{\mu}\right)  +\left(  m+gh|\phi|^{2}\right)  N\right]  =0\label{6}\\
\left(  \square+2h^{2}\left\vert \phi\right\vert ^{2}\right)  N-2\partial
_{\mu}Z^{\mu}\left(  m+gh|\phi|^{2}\right)  =0\label{7}\\
\partial_{\mu}F^{\mu\nu}+2mG^{\nu}=J^{\nu}\label{3}\\
\partial_{\mu}G^{\mu\nu}+\frac{m}{2}F^{\nu}=\frac{g}{2h}\varepsilon^{\mu
\kappa\nu}\partial_{\mu}J_{\kappa},\label{4}%
\end{gather}
where the current is $J_{\mu}=ih\left(  \phi^{\ast}\nabla_{\mu}\phi
-\phi\left(  \nabla_{\mu}\phi\right)  ^{\ast}\right)  $. We have three vector
fields, two of them coupled by a Chern-Simons term, and the other one coupled
to a scalar field. Despite this complicated mixing, Bogomoln'yi equations will
be help us to understand the role of each field in vortex formation.

Decoupling the Eqs. $\left(  \ref{3}\right)  $ and $\left(  \ref{4}\right)  $
from one another, we obtain%
\begin{align}
\left(  \square+m^{2}\right)  F^{\nu}  & =\varepsilon^{\mu\kappa\nu}%
\partial_{\mu}J_{\kappa}\left(  \frac{gm}{h}+1\right) \label{a1}\\
\left(  \square+m^{2}\right)  G^{\nu}  & =\left(  \square-\frac{hm}{g}\right)
\left(  \frac{-g}{2h}\right)  J^{\nu}.\label{a2}%
\end{align}

Using the critical coupling, $g=-\frac{h}{m}$, in the previous two equations
yields:%
\begin{gather}
\left(  \square+m^{2}\right)  F^{\nu}=0,\label{8}\\
G^{\nu}=\frac{1}{2m}J^{\nu}.\label{9}%
\end{gather}

From Eqs $\left(  \ref{8}\right)  $ and $\left(  \ref{9}\right)  ,$ we see
that the $A_{\mu}-$field decouples from the scalar field $\phi$. However, the
$G^{\mu}-$ field gives us%
\begin{equation}
\varepsilon^{\nu\alpha\beta}\partial_{\alpha}B_{\beta}=\frac{1}{2m}J^{\nu
}.\label{top}%
\end{equation}

The value of the critical coupling, $g=-\frac{h}{m},$ reveals the purely
topological character of the current, wich shall be a relevant information in
our analysis of the asymptotic behaviour of the field configurations.

\section{BP-states and asymptotic behaviour}

The explicit form of BPS-states can be worked out in a supersymmetric context.
This result is explained in ref. \cite{Hlousek:1991ic}. Based on that work, we
could define new supersimetric generators as follows,%
\begin{equation}
Q_{\pm}=Q_{\theta}\mp i\gamma^{0}Q_{\tau},\label{5}%
\end{equation}
where $Q_{\theta}$ e $Q_{\tau}$ are the initially generators for the $N=2$
supersymmetry. The generators $\left(  \ref{5}\right)  $ render manifest one
of the results of Houlsek and Spector \cite{Hlousek:1991ic},\
\begin{equation}
\left\{  Q_{+},\overline{Q}_{+}\right\}  =4\gamma^{0}\left(  P_{0}+Z\right)
;\ \left\{  Q_{-},\overline{Q}_{-}\right\}  =4\gamma^{0}\left(  P_{0}%
-Z\right)  .\
\end{equation}
where $Z$ it is a central charge of the extended supersymmetry.

Using these generators and setting to zero all fermionic variations, we can
obtain BPS-states; however, here, we shall present another approach (more
heuristic) that can be used also in the case of non supersymmetric models. To
do so, we begin with the energy density of our model
\begin{align}
E  & =\int d^{2}x\left\{  \frac{1}{2}\left(  \overrightarrow{E}^{2}%
+B^{2}\right)  +P\left(  \overrightarrow{G}^{2}+B^{2}\right)  +\right.
\nonumber\\
& +PS^{2}+e^{-2gM}\left(  D_{0}\varphi\right)  ^{\ast}\left(  D_{0}%
\varphi\right)  +\nonumber\\
& +e^{-2gM}\left(  D_{i}\varphi\right)  ^{\ast}\left(  D_{i}\varphi\right)
+P\left(  \partial_{0}M\right)  ^{2}+\nonumber\\
& +P\left(  \partial_{i}M\right)  ^{2}+\frac{1}{2}\left(  \partial
_{0}N\right)  ^{2}+\frac{1}{2}\left(  \partial_{i}N\right)  ^{2}+\nonumber\\
& \left.  +\left(  2mN+2gh\left\vert \phi\right\vert ^{2}\right)  S+U\right\}
,
\end{align}
where, contrary to the work of ref. \cite{Christiansen:1998gj}, the second
family of gauge potentials is not truncated. And this is one of ours
proposals: to understand the role of the $U\left(  1\right)  ^{\prime}$ factor
and its corresponding gauge potential, $B^{\mu}$, in the process of vortex formation.

Upon completion of squares,%
\begin{align}
E  & =\int d^{2}x\left\{  \frac{1}{2}\left[  B\mp h\left(  \frac{2m}%
{h}M-\left\vert \phi\right\vert ^{2}+v^{2}\right)  \right]  ^{2}+\right.
\nonumber\\
& +\frac{1}{2}\left(  E_{i}\pm\partial_{i}N\right)  ^{2}+P\left(  G_{0}\pm
S\right)  ^{2}+\nonumber\\
& +P\left(  G_{i}\pm\partial_{i}M\right)  ^{2}+e^{-2gM}\left\vert \left(
D_{0}\pm ihN\right)  \varphi\right\vert ^{2}+\nonumber\\
& +e^{-2gM}\left\vert \left(  D_{1}\pm iD_{2}\right)  \varphi\right\vert
^{2}\pm hB\left(  \frac{2m}{h}M-\left\vert \phi\right\vert ^{2}+v^{2}\right)
+\nonumber\\
& \mp E_{i}\partial_{i}N\mp2PG_{0}S\mp2PG_{i}\partial_{i}M\mp2e^{-2gM}%
NH_{0}+\nonumber\\
& \mp e^{-2gM}\left(  \frac{1}{h}\varepsilon_{ij}\partial_{i}H_{j}%
+hB\left\vert \varphi\right\vert ^{2}\right)  +\nonumber\\
& \left.  +\left(  2mN+2gh\left\vert \phi\right\vert ^{2}\right)  S+U\right\}
,
\end{align}
with%
\begin{equation}
H_{\mu}=-\frac{ih}{2}\left(  \varphi^{\ast}D_{\mu}\varphi-\varphi\left(
D_{\mu}\varphi\right)  ^{\ast}\right)  .
\end{equation}
Now, we drop all quadratic terms as we are interested in the minimum energy
configuration. Then, we obtain the BPS-equations:%

\begin{gather}
B\mp h\left(  \frac{2m}{h}M-\left\vert \phi\right\vert ^{2}+v^{2}\right)
=0;\ \ \ \\
\partial_{\mu}Z^{\mu}=S=\pm G_{0};\ \ \ E_{i}\pm\partial_{i}N=0;\label{2}\\
G_{i}\pm\partial_{i}M=0;\ \ \left(  \nabla_{1}\pm i\nabla_{2}\right)
\phi=0.\label{10}%
\end{gather}

Introducing Eq. $\left(  \ref{2}\right)  $ in $\left(  \ref{6}\right)  $ and
$\left(  \text{\ref{7}}\right)  $,$\ $we recover Eqs. $\left(  \ref{3}\right)
$ and $\left(  \text{\ref{4}}\right)  $, showing that BPS-states agree with
the results from the equations of motion, as expected. It is worthy to mention
that the field-strength for the Kalb-Ramond potential becomes the topological charge.

If asympotically we write $\phi=ve^{in\theta}$, then, from equation $\left(
\ref{10}\right)  $, we get%
\begin{gather}
\frac{1}{-i}\phi^{-1}\left(  \partial_{1}\pm i\partial_{2}\right)
\phi=-e^{\pm i\theta}\frac{n}{r},\\
-e^{\pm i\theta}\frac{n}{r}=e\left(  A_{1}\pm iA_{2}\right)  +g\left(
G_{1}\pm iG_{2}\right)  .
\end{gather}

Therefore, in the minimum energy configuration both fields, $A_{\mu}$ and
$G_{\mu}$, participate of the vortex formation. However, for the critical
coupling $\left(  g=-\frac{h}{m}\right)  $, asymptotically, only the field
that appears in the non-minimal coupling, $G_{\mu}$, is relevant for the
vortex configuration:%
\begin{equation}
2m\int d^{2}xb=Q_{top}=2m\Phi_{flux}.
\end{equation}

By analyzing the critical coupling and the asymptotic behaviour, we see that
the non-minimal coupling in the covariant derivative contributes directly to
the topological current, in agreement with equation $\left(  \ref{top}\right)
$.

\section{\bigskip A Possible Relation with a Condensed Matter Physics System}

The relation between a global vortex in the Abelian Higgs model and vortices
in a superfluid has been exploited in \cite{DS}. This work is developed in
$4D$ and basically two problems are found when we try to identify them. The
first difference has to do with the energy density that falls off like
$1/r^{2}$ in the case of the global vortex; on the other hand, vortices in a
superfluid have non-zero enegy density at infinity. The second main difference
is related with the angular momentum, that is well-defined for vortices in a
superfluid, but is zero for global vortices, when considering static configurations.

These problems have been solved when Davis and Shellard considered time
dependent equations and a non-trivial background, as below:%
\begin{equation}
G^{ijk}=\alpha\varepsilon^{ijk}.\label{ds}%
\end{equation}

This is clearly done in $4D$; but, it is similar to the equation $\left(
\ref{bc}\right)  $ that naturally shows up in $3D$. \ The reason why they
achieve this result is that $\left(  \ref{ds}\right)  $ simulates a
preferencial background for the superfluid and contributes with a non-zero
energy at infinity. An important fact to mention is that, in order to
introduce a non-trivial background in our $N=2-D=3$ model, the SSB must also
be realised by the Kalb-Ramond field. This has been done in $4D$ because the
scalar action and the Kalb-Ramond action are simply related by a canonical
transformation \cite{KE}. However in $3D$ the SSB cannot be realised by the
Kalb-Ramond field and will be entirely described by a scalar field.

Another relation of our $N=2-D=3$ with Condensed Matter concerns the gauge
action:
\begin{align}
S_{gauge}  & =\int d^{3}x\{-{\frac{1}{4}}F_{\mu\nu}F^{\mu\nu}%
+\,2\,m\,\varepsilon^{\mu\nu\alpha}A_{\mu}\partial_{\nu}B_{\alpha}\nonumber\\
& -{\frac{1}{2}}G_{\mu\nu}G^{\mu\nu}\}.\label{lag}%
\end{align}
In a lower-dimensional Condensed Matter system, the Chern-Simons-like term in
equation $\left(  \ref{lag}\right)  $ could also provide a non-trivial
background. This mixing has been studied as an effective theory \cite{DJ} in
which a dynamical vortex is coupled with a superfluid film at zero
temperature. In the $\varepsilon^{\mu\nu\alpha}A_{\mu}\partial_{\nu}B_{\alpha
}$-term, the $A_{\mu}-$ field is chosen as the responsible for the vortex
formation and the$B_{\mu}-$ field as the electromagnetic potential, which
becomes part of the source that describes a uniform magnetic field. Also here,
time-dependent equations must be considered.

Therefore, references \cite{DS} and \cite{DJ} seem to be very closely related
with our $N=2-D=3$ supersymmetric model. We have pointed it out as a
suggestion for future works. In both cases, the requirement of considering
dynamical solutions is evident, so time dependence must be considered in
equations $\left(  \ref{a1}\ \text{and\ \ref{a2}}\right)  $.

\section{General Conclusions}

In this work, we have shown that the Kalb-Ramond current has a topological
conservation law in four dimensions. So, it seems reasonable that the coupling
of the KR field to any other theory must be non-minimal. This also supports
the non-existence of a non-Abelian generalization for these theories. Our
result agrees with the "no-go"\ theorem \cite{Henneaux:1997mf}. We however
would like to point out the efforts in building up an interesting extension of
the gauge approach to allow minimal couplings of the 2-form gauge potential
\cite{botta1}, \cite{botta2}.

In the study of vortex formation, the KR-field strength in $1+2$ dimensions is
a simple constant and it couples to the present model as the topological
charge of the vortex. This may also describe a non-trivial background. Also
the non-minimal coupling of the vector field in the covariant derivative
becomes directly identified with the topological current, which seems to
stabilize the topological solutions for configuration of non-minimal energy.

We analyzed how BPS-states in this model reduce the number of differential
equations and give us some insight on the role of each field whenever half of
the supersymmetry charges become zero. We see that the mixing of the minimal
and non-minimal couplings contributes for the ansatz on the scalar field, in
general. However, with the critical coupling, $g=-\frac{h}{m}$, only the
non-minimal coupling is actually relevant for the vortex configuration.

Finally, our perspectives are to study the possibility of having a minimal
coupling of the KR model in higher dimensions and study whether or not this
coupling is allowed in presence of a gravity background. It would also be
interesting to explore further the relation beetwen our $N=2-D=3$ and
dynamical vortices in a superfluid film.

\textbf{Acknowledgments:}

The authors express their gratitude to A.L.M.A. Nogueira for clarifying
discussions and very helpful suggestions. They also thank CNPq-Brasil for the
financial support.

\end{document}